\documentclass[nohyperref]{article}

\usepackage{microtype}
\usepackage{graphicx}
\usepackage{subfigure}
\usepackage{booktabs} 

\usepackage{hyperref}

\usepackage[accepted]{icml2023}

\usepackage{amsmath}
\usepackage{amssymb}
\usepackage{mathtools}
\usepackage{amsthm}
\usepackage{makecell} 

\usepackage[capitalize,noabbrev]{cleveref}

\theoremstyle{plain}

\theoremstyle{definition}

\theoremstyle{remark}

\usepackage[textsize=tiny]{todonotes}

\icmltitlerunning{A normative theory of social conflict}

\begin{document}

\twocolumn[
\icmltitle{A normative theory of social conflict}




\begin{icmlauthorlist}
\icmlauthor{Sergey Shuvaev}{cshl}
\icmlauthor{Evgeny Amelchenko}{sbu}
\icmlauthor{Dmitry Smagin}{icg}
\icmlauthor{Natalia Kudryavtseva}{icg}
\icmlauthor{Grigori Enikolopov}{sbu}
\icmlauthor{Alexei Koulakov}{cshl}
\end{icmlauthorlist}

\icmlaffiliation{cshl}{Cold Spring Harbor Laboratory, Cold Spring Harbor, NY, USA}
\icmlaffiliation{sbu}{Center for Developmental Genetics, Stony Brook University, Stony Brook, NY, USA}
\icmlaffiliation{icg}{Institute of Cytology and Genetics, Russian Academy of Sciences (Siberian Branch), Novosibirsk, Russia}

\icmlcorrespondingauthor{Alexei Koulakov}{koulakov@cshl.edu}

\icmlkeywords{Machine Learning, ICML}

\vskip 0.3in
]



\printAffiliationsAndNotice{}  

\begin{abstract}
Social hierarchy in animal groups carries a crucial adaptive function by reducing conflict and injury while protecting valuable group resources. Social hierarchy is dynamic and can be altered by social conflict, agonistic interactions, and aggression. Understanding social conflict and aggressive behavior is of profound importance to our society and welfare. In this study, we developed a quantitative theory of social conflict. We modeled individual agonistic interactions as a normal-form game between two agents. We assumed that the agents use Bayesian inference to update their beliefs about their strength or their opponent’s strength and to derive optimal actions. We compared the results of our model to behavioral and whole-brain neural activity data obtained for a large population of mice engaged in agonistic interactions. We find that both types of data are consistent with the first-level Theory of Mind model (1-ToM) in which mice form both ``primary'' beliefs about their and their opponent’s strengths as well as the ``secondary'' beliefs about the beliefs of their opponents. Our model helps identify brain regions that carry information about these levels of beliefs. Overall, we both propose a model to describe agonistic interactions and support our quantitative results with behavioral and neural activity data. 
\end{abstract}

\section{Introduction}
Social hierarchy has an important adaptive role in animals ranging from insects to primates. The formation of a social hierarchy helps to mold the group’s structure, ensuring a degree of flexibility in changing circumstances. In primates and rodents, social conflict and, in particular, inter-male aggression, play a critical role in both retaining and altering the social structure of the group. Although the overall role of social conflicts is crucial, e.g., enabling the allocation of limited resources via a few-shot formation of social hierarchies \cite{scott1971, rosell2015, chester2016}, unjustified social conflicts in resource-rich environments may be maladaptive and lead to drastic negative consequences \cite{neumann2010, chester2016, golden2019}. Excessive or pathological male aggression is one of the most destructive forces in human society. Despite the continued experimental studies of social conflict and the underlying neural circuitry, its theoretical framework and quantitative principles remain to be understood. Here, we develop a theoretical model for social agonistic interactions and use experimental data obtained in male mice in a comprehensive paradigm of chronic conflict to validate our model.

\section{Related work}
\subsection{Behavioral biology of aggression}

Aggressive behaviors and social hierarchy have been extensively studied in humans, other primates, and rodents. Studies in mice -- a model organism whose social status can be effectively manipulated by experimental and genetic means, have provided a wealth of knowledge about the formation, maintenance, and plasticity of the social hierarchy, aggression and defeat, and dominant and subordinate status \cite{wong2016, hashikawa2017}. These studies often use variations of the chronic social conflict paradigm, where mice are allowed to engage in agonistic interactions for a limited time on a daily basis \cite{kudryavtseva2000, miczek2001, golden2019}. The chronic social conflict paradigm has uncovered key features of aggressive behavior including the potentiating effect of repeated victory, the aversive effects of repeated defeat, and the similarities between pathological aggression and drug addiction \cite{miczek2001, aleyasin2018, golden2019}. Here, we use a comprehensive paradigm of chronic social conflict to generate distinct social statuses in mice and propose a normative theory to explain related aggressive behavior. Because the studies of aggression convincingly demonstrate the evolutionary preservation of its basic mechanisms \cite{wang2010, watanabe2017}, our results may be relevant to human behavior. 

\subsection{Game theory}

Optimal behaviors of interacting agents are conventionally described in terms of game theory. The game theory considers rational agents developing their strategies in order to maximize rewards. The rewards received by the agents depend on their actions and the actions of their opponents. The acquisition of optimal strategies in games can be described by probabilities of available actions \cite{smith1982, cressman2003}. Such strategies of agents co-evolve to reinforce higher-reward actions until the rewards can't grow any longer (Nash equilibrium). Game-theoretical approaches have been used in models of human and animal behaviors in multi-agent settings including agonistic interactions \cite{smith1974, hofbauer1998, wilson2000, lorenz2005}. Here, we use game theory to model agonistic interactions in the condition of chronic social conflict in mice. 

\subsection{Beliefs and Theories of Mind}

To accumulate evidence in partially observed environments humans and animals may maintain their probabilistic internal models of the environment -- the ``beliefs'' -- based on which their actions can be viewed as rational, maximizing a reward function \cite{fahlman1983, alefantis2021}. The agents' rewards and beliefs can be inferred from their behavior using inverse control techniques, which maximize the likelihood of the observed behavior based on a particular hidden dynamics model \cite{russell1998, choi2011, dvijotham2010, kwon2020}. In biologically relevant multi-agent settings, beliefs are studied in the Theories of Mind (ToM) framework which proposes that human and animal agents may maintain beliefs about the beliefs of their adversaries \cite{baker2011} or aides \cite{khalvati2019}. As previous studies were successful in inferring beliefs \cite{schmitt2017, alefantis2021} and regressing them to neural activity in simulations \cite{wu2020} and low-resolution fMRI imaging \cite{koster2013}, here we propose a way to infer task-related beliefs in mice and compare them to high-resolution data of the whole-brain neural activity.

\subsection{C-Fos as a whole-brain marker of neural activity}

The search for brain regions accumulating evidence about the environment requires large-scale neural activity data. Such data can be obtained by monitoring the levels of c-Fos, an immediate early gene whose activation reflects neuronal activity \cite{sagar1988, herrera1996}. The c-Fos data lacks temporal resolution, yet it allows observing whole-brain activity at high spatial resolution without using equipment that may affect animals' choices. Local expression of c-Fos has implicated certain brain regions in agonistic interactions \cite{hashikawa2017, aleyasin2018, diaz2020, wei2021}. Here, we use 3D light-sheet microscopy of c-Fos in whole-brain samples \cite{renier2016} to identify brain-wide neural activity in animals with varying exposure to social conflict. We compare the c-Fos data to beliefs identified based on behavior in individual mice and report the regions which may be involved in the computation of conflict-related variables in the brain.

\section{Results: normative model of social conflict}

The goal of this work is to build a quantitative theory for the formation of social conflict-related behavioral states in mice. In \cref{met:behavior} we describe a mouse behavioral paradigm where we recorded the actions leading to different behavioral states and the brain activities corresponding to these states. In \cref{met:gametheory,met:beliefs,met:bayesian} we define the model of social conflict in which game-theory optimal actions of agents rely on beliefs about their strength. In \cref{res:results} we examine hypotheses about the reward schedule, information availability, and evidence accumulation related to social conflict. We compare our results to behavioral data in \cref{res:behavioral} and to neural data in \cref{res:neural}. We discuss our findings in \cref{res:theory}.
\begin{figure}[ht]
\vskip 0.2in
\begin{center}
\centerline{\includegraphics[width=\columnwidth]{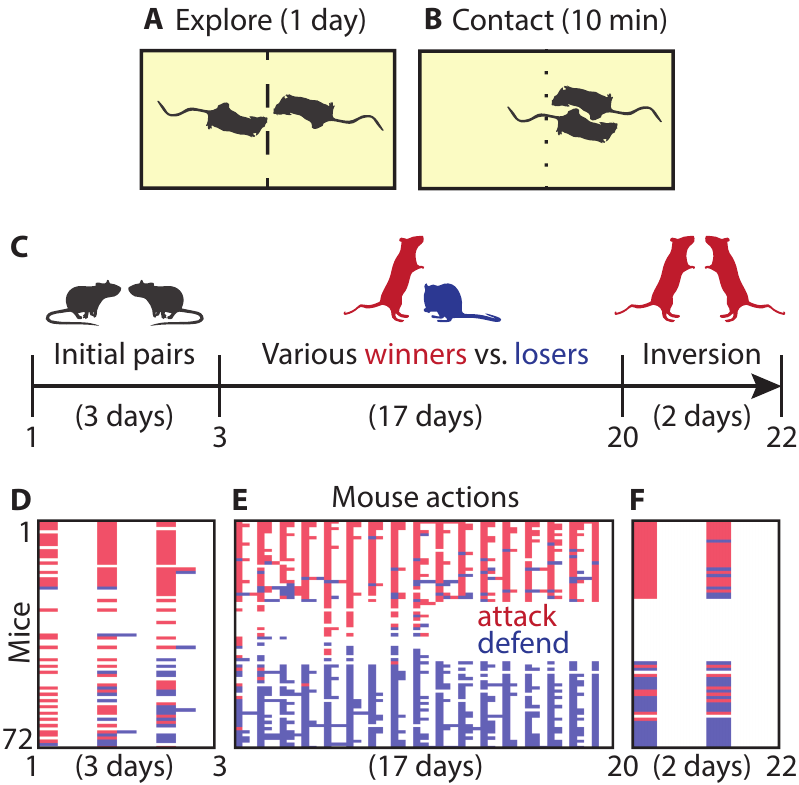}}
\caption{The chronic social conflict paradigm. (A-B) Stages of a single sensory contact event. (C) Stages of a multi-day experiment. (D-F) Recorded actions of mice in the experiment.}
\label{fig:setting}
\end{center}
\vskip -0.2in
\end{figure}

\subsection{Chronic social conflict paradigm}
\label{met:behavior}

To observe animals with varied behavioral states we implemented the chronic social conflict paradigm \citep{kudryavtseva2020} (\cref{app:behavior}) as follows. Pairs of weight-matched (as a proxy for being strength-matched) mice were placed in cages separated by a perforated partition (\cref{fig:setting}A). Once daily, the partition was removed for 10 minutes to enable agonistic interactions between mice (\cref{fig:setting}B). Although the majority of the mice displayed aggressive behavior upon first interactions, the dominance relationships have formed in most of the pairs after 2-3 aggressive encounters (2-3 days; \cref{fig:setting}C,D). Afterward, each winning mouse remained in its cage, while each losing mouse was daily relocated to an unfamiliar cage with an unfamiliar winning mouse (\cref{fig:setting}C). Regardless of no longer being weight-matched, mice have remained in their dominant or submissive behavioral states, transitioning to the maladaptive regime of social conflict-related decision-making (\cref{fig:setting}E). After 20 days of interactions, mice were exposed to opponents of equal behavioral state (\cref{fig:setting}C). The newly formed pairs underwent two more days of agonistic interactions throughout which new dominance relationships were established (\cref{fig:setting}F). 

In the experiments, we tested multiple groups of mice subjected to different numbers of agonistic interactions before performing the whole-brain imaging of c-Fos expression as a proxy for neuronal activation. 17 mice participated in the experiment for 3 days forming the groups of ``winners'' $W_{3}$ and ``losers'' $L_{3}$. Another batch of 16 mice participated for 10 days and 49 mice participated for 20 days, similarly forming the groups $W_{10}$, $L_{10}$, $W_{20}$, and $L_{20}$. 52 mice participated for 22 days forming the groups of ``winners-remain-winners'' ($WW$), ``winners-become-losers'' ($WL$), ``losers-become-winners'' ($LW$), and ``losers-remain-losers'' ($LL$). Thus, a total of 134 mice participated in the experiment, leading to the behavioral data on their opponents, actions, and agonistic interaction outcomes; the c-Fos expression was analyzed in a total of 54 mice from all the experimental conditions and 18 control animals.

\subsection{Game-theoretical model of social conflict}
\label{met:gametheory}

Below, we build a model based on the chronic social conflict paradigm. In this section, we start from an approximation in which each agent has all information about itself and its opponents. For that case, we define the optimal actions for each agent using the game theory.

We formalize our behavioral paradigm as a normal form game (\cref{app:gametheory}), i.e. a process in which, in each iteration, two agents have to decide simultaneously what action $a$ to take. We defined the possible actions $a$ as ``attack'' or ``defend''. Depending on the actions $a_1$ and $a_2$ selected by the two agents respectively, they received rewards $r$ defined as follows. If both agents chose to “defend”, no fight happened, leading to a zero reward $r_1 = r_2 = 0$ assigned to each agent. If both agents ``attacked'', the outcome of the game was defined by their strengths $s$, an additional parameter assigned to each agent in the model. The outcome probability $p^{win}$ was defined by the softmax rule over the strengths parameterized with the ``outcome confidence'' $\beta_o$:
\begin{equation}
p_i^{win} = Z^{-1}\exp(\beta_o s_i).
\end{equation}
Here and below $Z$ denotes the normalization coefficient. Once the outcome was determined, the winning agent received a reward of $r = +1$, and the losing agent expended a cost of $r = -\mathcal{A}$. The reward expectation was equal to:
\begin{equation}
\kappa_i = 1 \cdot p_i^{win} + (-\mathcal{A}) \cdot (1 - p_i^{win}) = (1 + \mathcal{A}) p_i^{win} - \mathcal{A}.
\end{equation}

\begin{figure}[ht]
\vskip 0.1in
\begin{center}
\centerline{\includegraphics[width=\columnwidth]{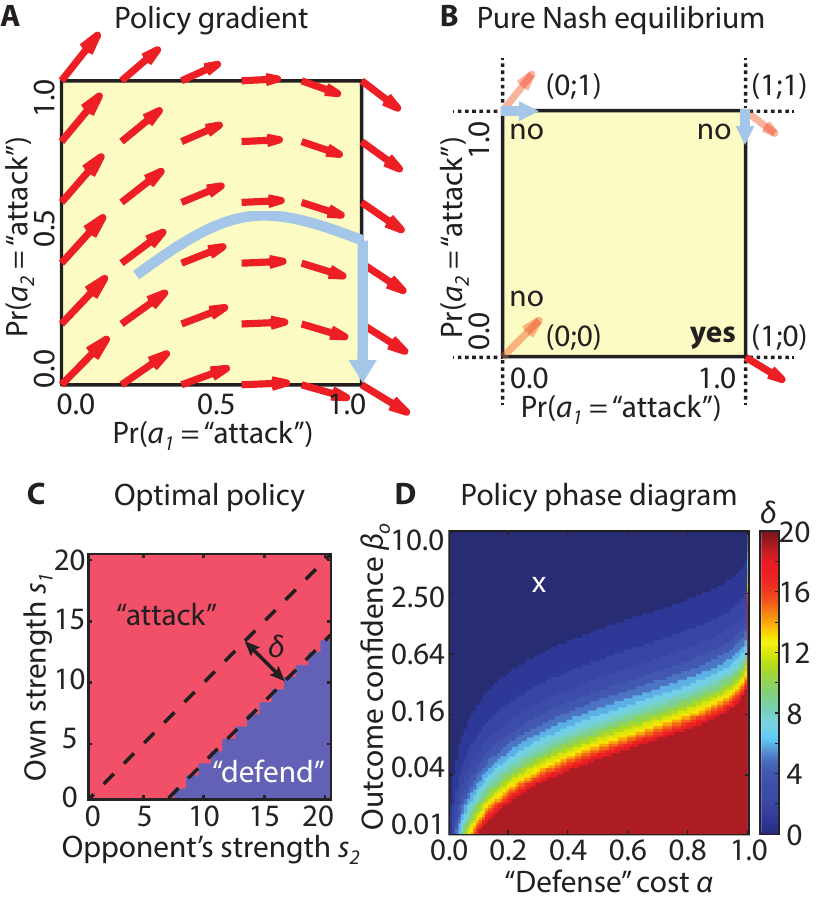}}
\caption{Game-theoretical model of the chronic social conflict paradigm. (A) Policy gradients. (B) Gradient orientations yield a pure strategy as a Nash equilibrium. (C) Optimal policy and its dependence on relative strength. (D) Dependence of optimal policy on the model parameters.}
\label{fig:gametheory}
\end{center}
\vskip -0.2in
\end{figure}

The cost of loss was reduced if one of the agents chose to ``defend'' while the other agent ``attacked''. In that case, the ``attacking'' agent always won and received the reward of $r = +1$ while the losing agent expended the cost of $r = -\alpha$ with $\alpha < \mathcal{A}$. The reward expectations were described in the payoff matrix $\hat{R}_i$ whose rows correspond to the actions of the agent (``attack'' and ``defend'') and columns correspond to the actions of its opponent:
\begin{equation}
\hat{R}_i = \begin{pmatrix} \kappa_i & 1 \\ -\alpha & 0 \end{pmatrix}.
\end{equation}
To determine optimal strategies in this game, we used evolutionary game theory. In this approach, the goal of every participant was to maximize its expected reward $\mathbb{E}[r_i]$:
\begin{equation}
\mathbb{E}[r_1] = P_1^T \hat{R}_1 P_2.
\end{equation}
Here the vectors $P_i$ define the probabilities to ``attack'', $p_i$, and to ``defend'', $1 - p_i$, for the agent number $i$:
\begin{equation}
P_i \equiv \begin{pmatrix} p_i \\ 1 - p_i \end{pmatrix}.
\end{equation}
A similar expression can be written for the expected reward of the second agent $\mathbb{E}[r_2]$. To maximize the rewards $\mathbb{E}[r_i]$, we computed their gradients with respect to probabilities to ``attack'' $p_i$ (an agent could only update its own policy, but not that of the opponent). We used these gradients to update the policies leading to joint maximization of the expected rewards (\cref{fig:gametheory}A, \cref{sup:gametheory}):
\begin{equation}
\begin{cases} \dot{p}_1 \propto (\kappa_1 + \alpha - 1) p_2 + 1; \\ \dot{p}_2 \propto (\kappa_2 + \alpha - 1) p_1 + 1.\end{cases}
\end{equation}
The probabilities of actions can be only defined in the range $0 \leq \{p_1, p_2\} \leq 1$. Their evolution, governed by the equations above (red arrows in \cref{fig:gametheory}A), may converge to a fixed point within this range, forming a ``mixed'' strategy. Alternatively, the probabilities of actions may converge to zeros or ones (blue arrows in \cref{fig:gametheory}A) forming ``pure'' strategies. To determine optimal ``pure'' strategies, we chose the strategies whose reward gradients (red arrows in \cref{fig:gametheory}B) pointed outwards the $[0 - 1]$ interval for both agents.

We represented the optimal policies via the tensor $A$ of probabilities for each possible action $a$ depending on the agent's strength $s_1$ and their opponent's strength $s_2$, averaging over all optimal ``pure'' and ``mixed'' strategies (\cref{fig:gametheory}C):
\begin{equation}
A_{aij} = Pr(a_1 = a | s_1 = i, s_2 = j).
\end{equation}

The resulting optimal policies in the model depended on the relative strengths of the agents (\cref{fig:gametheory}C). We parameterized the optimal policies with the maximum relative strength of the agents $\delta = s_2 - s_1$ at which it was still optimal to ``attack''. We then explored how $\delta$ depends on the task parameters $\alpha$ and $\beta_o$ and found that most of these parameters' values correspond to the optimal strategy with $\delta = 0$ (``x'' in \cref{fig:gametheory}D), i.e. to ``attacking'' any opponent who is weaker or equal. Overall, the acquisition of the optimal policies for the agents in our game is summarized in \cref{alg:gametheory}.

\subsection{Partial observability: beliefs about strength}
\label{met:beliefs}

The game theory predicts a static policy, unchanging over time. Conversely, weight-matched mice in the experiment initially attacked each other but later split into always-attacking ``winners'' and ever-defending ``losers'' (\cref{fig:setting}E). To account for this dynamic, we expand our model to a scenario where agents do not possess perfect information about their strengths but accumulate this information over the task. In this section, we model the agents' information about strengths via beliefs -- the probability distributions for an agent to belong to a certain strength category. We use the Bayes rule to initialize the beliefs about the animals' strengths using information about their body weight (\cref{app:beliefinit}) as follows.

We used body weights $w$ of the animals as a proxy of their strengths $s$ \cite{andersson1996, cooper2020}:
\begin{equation}
w_i \propto s_i.
\end{equation}
We modeled the animals' initial estimates of their body weights $\tilde{w}$ as normally distributed around true values $w$.
Under this assumption, we reconstructed the probability distributions for animals' strengths (their initial beliefs) $B^{i} \equiv Pr(s = i | \tilde{w})$ using their body weights $\tilde{w}$ in the Bayes rule:
\begin{equation}
\begin{gathered}
B^{i} \equiv Pr(s = i | \tilde{w}) = \frac{P_{\mathcal{N}(i, \sigma)}(\tilde{w}) Pr(s = i)}{Pr(\tilde{w})},
\end{gathered}
\end{equation}
where $P_{\mathcal{N}(i, \sigma)}(\tilde{w})$ is a normal distribution probability density function representing the noise in the estimate of the animal's weight; $Pr(\tilde{w})$ is the distribution of the animals' estimated weights $\{\tilde{w}_i\}$, and $Pr(s)$ is the distribution of their strengths $\{s_i\}$. To denoise the strength distribution $Pr(s)$, we approximated the experimentally observed weight distribution $Pr(w) \propto Pr(s)$ with a normal distribution.

We consider up to four types of beliefs. The two ``primary'' beliefs describe the animals' estimates of their own strength and of the strength of the opponent. The two ``secondary'' beliefs estimate the ``primary'' beliefs of the opponent. The animals were expected to better estimate their own strength compared to that of their opponents. To this end, we used separate uncertainty parameters for estimating one's own strength ($\sigma = \sigma_1$) and that of an opponent ($\sigma = \sigma_2$). For the ``secondary'' beliefs, the uncertainties were combined ($\sigma = \sigma_1 + \sigma_2$). Overall, initializing the beliefs about the animals' strengths is summarized in \cref{alg:beliefinit}.

\subsection{Evidence accumulation: Bayesian update of beliefs}
\label{met:bayesian}

\begin{figure}[ht]
\vskip 0.1in
\begin{center}
\centerline{\includegraphics[width=\columnwidth]{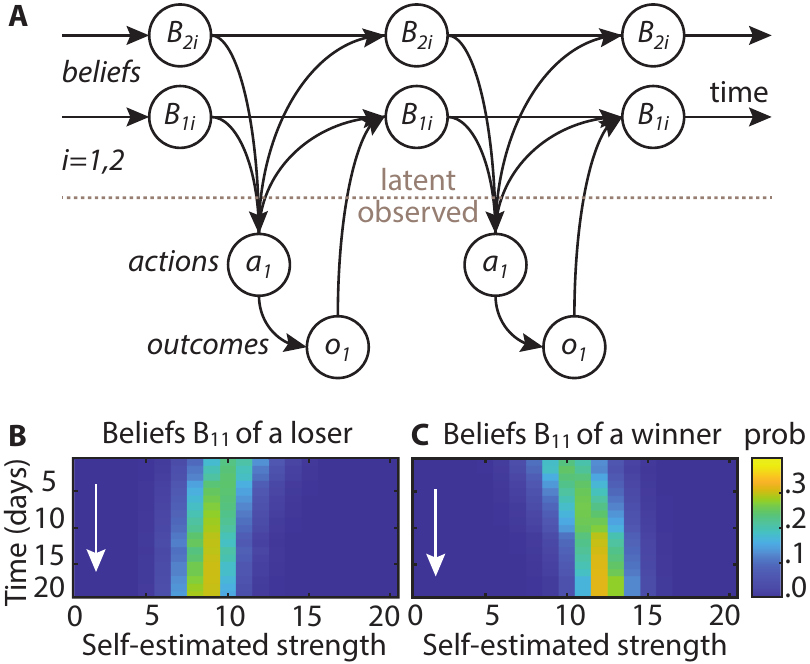}}
\caption{Bayesian update of beliefs in the model. (A) Belief update diagram for two mice. (B,C) Reconstructed belief dynamics for a representative loser and winner mice.}
\label{fig:bayesian}
\end{center}
\vskip -0.2in
\end{figure}

In this section, we expand our model with an update mechanism for the agents' beliefs (\cref{app:beliefupdate}). We update the beliefs $B$ using the Bayes rule and the information about actions $a$ and outcomes $o$ of agonistic interactions (\cref{fig:bayesian}A).

We define the \textit{outcome tensor} $O^{oab}_{ij}$ with the probabilities of the outcome $o_1 = o$ for an agent of the strength $s_1 = i$ after choosing an action $a_1 = a$ provided that the opponent has the strength $s_2 = j$ and chose the action $a_2 = b$: 
\begin{equation}
O^{oab}_{ij} = Pr(o_1 = o | a_1 = a, a_2 = b, s_1 = i, s_2 = j).
\end{equation}
The \textit{reward tensor} $R^{ab}_{ij}$ describes the expected reward $\mathbb{E}[r_1]$ for an agent of the strength $s_1 = i$ after choosing an action $a_1 = a$ provided that the opponent has the strength $s_2 = j$ and chose the action $a_2 = b$. The reward expectation was based on probabilities of possible outcomes:
\begin{equation}
R^{ab}_{ij} = \sum_o r(o, a, b) O^{oab}_{ij}.
\end{equation}

The \textit{outcome-action} and \textit{reward-action tensors} predicted the probability of outcome $o_1$ and expectation of reward $r_1$ assuming that the opponent's actions $a_2$ are optimal:
\begin{equation}
[RA]^{a}_{ijkl} \equiv \sum_b R^{ab}_{ij} \textrm{conv}_{kl}(A, P_{\mathcal{N}(0, \sigma_1 + \sigma_2)})^{b}_{kl};
\end{equation}
\begin{equation}
[OA]^{oab}_{ijkl} \equiv O^{oab}_{ij} \textrm{conv}_{kl}(A, P_{\mathcal{N}(0, \sigma_1 + \sigma_2)})^{b}_{kl}.
\end{equation}
Here the opponent's action is estimated based on the ``secondary" beliefs and the estimated outcome is based on the ``primary" beliefs. The convolution reflects that the opponent's action is based on the distributional belief rather than on a point estimate of the strengths. The standard deviation $\sigma_1 + \sigma_2$ applies both to the agent's estimate of the opponent's strength and to the estimated opponent's estimate of the agent's strength.

To decide on an action $a_1$, an agent maximized its reward $r_1$. We computed this reward by summing the reward-action tensor $[RA]^{a}_{ijkl}$ multiplied by the belief tensors $B_{11}^i$, $B_{12}^j, B_{21}^k$, $B_{22}^l$ reflecting probability distributions for strengths:
\begin{equation}
a_1 = \mathop{\textrm{softmax}}_{a} \sum_{ijkl} [RA]^{a}_{ijkl} B_{11}^i B_{12}^j B_{21}^k B_{22}^l.
\end{equation}
Here the belief tensors $B_{11}$ and $B_{12}$ describe the ``primary'' beliefs of an agent about its own strength and the strength of its opponent respectively. Likewise, the belief tensors $B_{21}$ and $B_{22}$ describe the ``secondary'' beliefs, i.e. the agent's estimate of its opponent's ``primary'' beliefs. The indices $i,j,k,l$ iterate over particular strengths, e.g. $B_{11}^5 = 0.2$ would mean that, according to the agent's belief, the probability of its own strength to be equal to 5 constitutes 20\%. To update the agents' beliefs using the observed actions and outcomes $\{o_1, a_2\}$, we used the Bayes rule individually for every element of each belief tensor $B_{**}^{i}$:
\begin{equation}
\begin{gathered}
Pr(s_1 = i | \{o_1, a_2\}) = \\
 = \frac{Pr(\{o_1, a_2\} | s_1 = i) Pr(s_1 = i)}{Pr(\{o_1, a_2\})}.
\end{gathered}
\end{equation}
The probability of an outcome $o_1$ and an action $a_2$ for the agent of the strength $s_1 = i$ was derived from the outcome-action tensor $[OA]^{o_1 a_1 a_2}_{ijkl}$:
\begin{equation}
Pr(\{o_1, a_2\} | s_1 = i) = \sum_{jkl} [OA]^{o_1 a_1 a_2}_{ijkl}.
\end{equation}
The probability for the agent to be of strength $s_1 = i$ was taken from the belief tensor $B_{11}^{i}$:
\begin{equation}
    Pr(s_1 = i) = B_{11}^{i}.
\end{equation}
The marginal probability of the observation $\{o_1, a_2\}$ was defined by the outcome-action tensor $[OA]^{o_1 a_1 a_2}_{ijkl}$ scaled by the beliefs about strength $B_{11}^i$, $B_{12}^j$, etc. to weigh optimal actions with probabilities of their underlying strengths:
\begin{equation}
\begin{gathered}
Pr(\{o_1, a_2\}) = Pr(o_1 | a_2) Pr(a_2) = \\
= \sum_{ijkl} [OA]^{o_1 a_1 a_2}_{ijkl} B_{11}^i B_{12}^j B_{21}^k B_{22}^l.
\end{gathered}
\end{equation}
Together, the four equations above formed the update rule for agent beliefs based on their observations:
\begin{equation}
Pr(s_1 = i | \{o_1, a_2\}) = \frac{\sum_{jkl} [OA]^{o_1 a_1 a_2}_{ijkl} B_{11}^i B_{12}^j B_{21}^k B_{22}^l} {\sum_{ijkl} [OA]^{o_1 a_1 a_2}_{ijkl} B_{11}^i B_{12}^j B_{21}^k B_{22}^l}.
\end{equation}
The update could proceed at an arbitrary learning rate $\varepsilon$:
\begin{equation}
    B_{11}^i \leftarrow \varepsilon Pr(s_1 = i | \{o_1, a_2\}) + (1 - \varepsilon) B_{11}^i
\end{equation}
Overall, the update procedure for the beliefs about the animals' strengths is summarized in \cref{alg:beliefupdate}.

\section{Results: mechanisms of animal behavior}
\label{res:results}
\subsection{Model fit and comparison}
\label{res:behavioral}

We used the framework defined above to test several hypotheses about mouse social conflict-related choices. To test the hypotheses against alternatives, we optimized pairs of models on the training data (52 mice participating for 22 days). To this end, we specified the negative log-likelihood (NLL) function $\mathcal{L}$ comparing the action probabilities predicted in our model to mouse actions logged in the experiment:
\begin{equation}
\mathcal{L} = -\sum_{mt}(\log Pr(a_m^t) + \log Pr(o_m^t)).
\end{equation}
To fit the model parameters, we minimized the NLL regularized with the $l_2$ norm of its arguments (\cref{app:modelfit}). We chose the regularization coefficient using fits on a simulated experiment (\cref{sup:modelfit}, \cref{fig:parameter_inference}C-H). For the real mice in the experiment, we used the data spanning all 22 days to propagate the beliefs but the NLL was only computed for the data from days 1-3 and 21-22 to avoid the impact of repeated actions on days 4-20 (\cref{alg:parameterinference}, \cref{sup:modelfit}, \cref{fig:parameter_inference}A-B). To compare the models, we computed the changes in the NLLs based on the predictions of the models for the testing data (82 mice participating for 3/10/20 days; \cref{app:modelcompare}). We performed the t-test on the changes in NLLs for individual mice followed by the false discovery rate (FDR) correction. Our comparisons were iterative: they lasted until a model under consideration outperformed all the other models. Below, we report the results for the final round of these comparisons. The implementation details are described in \cref{app:modelcompare}. Training set results are reported in \cref{sup:model_comparison} and \cref{fig:model_comparison_training}

\subsubsection{Baseline models}

\begin{figure}[ht]
\vskip 0.1in
\begin{center}
\centerline{\includegraphics[width=\columnwidth]{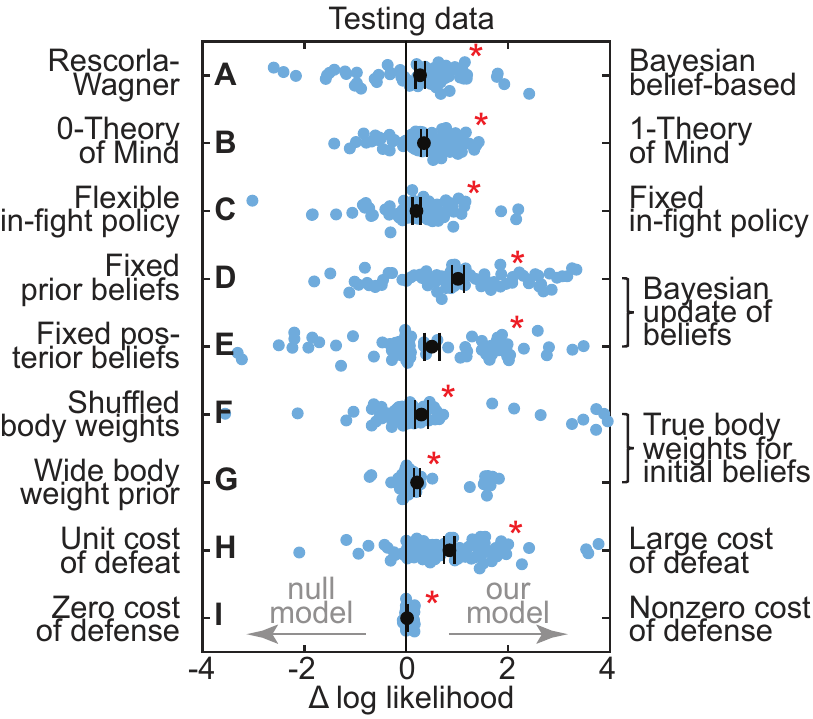}}
\caption{Model comparison on testing data. Here (*) indicates a significant difference between the models (t-test $p \leq 0.05$), (ns) indicates a non-significant difference (t-test $p > 0.05$), and the whiskers show the mean $\pm$ the standard error of the mean.}
\label{fig:model_comparison}
\end{center}
\vskip -0.2in
\end{figure}

Here we use the conventional approach and compare our model with the usual baseline models \cite{devaine2014, khalvati2019}. The comparison results are displayed in \cref{fig:model_comparison}; below we report the mean values for the differences in NLLs and the corresponding standard errors of the means. We used the FDR threshold $q = 0.05$ to evaluate the significance of tested hypotheses; the significant differences are marked with the $^*$ sign; the insignificant ones are marked with the $^{ns}$ sign both in the text and in the figure.

To determine the \textbf{class of algorithms} used by the animals in our task, we compared the Bayesian belief-based model with the Rescorla-Wagner reinforcement learning model where each action (“attack”, “defend”) is associated with a value updated based on the rewards (\cref{fig:model_comparison}A). Our tests suggest that the mouse actions in the experiment were more consistent with the Bayesian update of beliefs ($\Delta \textrm{NLL}_{train} = 0.30 \pm 0.15^*$; $\Delta \textrm{NLL}_{test} = 0.27 \pm 0.09^*$).

To analyze the \textbf{depth of reasoning} consistent with the animals’ actions, we compared the Bayesian belief-based model where the decisions were based on the ``primary'' beliefs and the game theory optimum (a zeroth-order Theory of Mind, 0-ToM) with a model including the ``secondary'' beliefs to predict the opponents' actions and choose the best response to these (1-ToM) (\cref{fig:model_comparison}B). We found that the animals’ actions in our experiment are better described with the 1-ToM model computing both ``primary'' and ``secondary'' beliefs ($\Delta \textrm{NLL}_{train} = 0.33 \pm 0.09^*$; $\Delta \textrm{NLL}_{test} = 0.35 \pm 0.06^*$).

To assess the \textbf{flexibility of policy} during a single agonistic interaction, we compared a model where animals decide on their actions before the interaction (“fixed policy”) with the model where animals adapt their actions to the opponent’s actions, thus converging to a joint equilibrium (“flexible policy”) (\cref{fig:model_comparison}C, \cref{sup:gametheory_flex}). This equilibrium may differ from the Nash equilibrium in the “fixed policy” as the animals’ beliefs may not be symmetric. We found that the animals’ actions were better described with the “fixed policy”, unchanging within a single agonistic interaction but evolving between interactions during chronic social conflict ($\Delta \textrm{NLL}_{train} = 0.50 \pm 0.12^*$; $\Delta \textrm{NLL}_{test} = 0.20 \pm 0.08^*$).

\subsubsection{Ablation studies}

To infer the \textbf{dynamics of beliefs} that best describe the animals’ actions, we compared the Bayesian belief-based model (“dynamic beliefs”) with the models with fixed beliefs (“static beliefs”; equivalent to a zero learning rate) \cite{baker2011}. We considered two types of “static beliefs”: “prior beliefs”, identical to those used to initialize the “dynamic beliefs” model (\cref{fig:model_comparison}D), and “posterior beliefs”, identical to the output of the “dynamic beliefs” model reflecting the most complete knowledge about the animals’ strengths (\cref{fig:model_comparison}E). We found that the “dynamic beliefs” model has better explained the animal data (``prior beliefs'': $\Delta \textrm{NLL}_{train} = 0.95 \pm 0.17^*$; $\Delta \textrm{NLL}_{test} = 1.02 \pm 0.12^*$; ``posterior beliefs'': $\Delta \textrm{NLL}_{train} = 0.50 \pm 0.12^*$; $\Delta \textrm{NLL}_{test} = 0.51 \pm 0.15^*$). We did not perform an additional ablation study for the learning rate coefficient in the ``dynamic beliefs'' model as, according to the fits, it was equal to one.

To test the \textbf{role of body weight} in the initial beliefs about animals’ strengths, we compared the model where the beliefs were initialized using the animals’ body weights with the model where the body weights were shuffled (\cref{fig:model_comparison}F) and with the model where the correct weights were used but their prior distribution was wider than the true distribution (\cref{fig:model_comparison}G). Our comparison shows that using the true animals’ body weights and their distribution to initialize beliefs has positively affected the predictions (``shuffled weights'': $\Delta \textrm{NLL}_{train} = 0.10 \pm 0.07^{ns}$; $\Delta \textrm{NLL}_{test} = 0.30 \pm 0.13^*$; ``wider prior'': $\Delta \textrm{NLL}_{train} = 0.08 \pm 0.04^*$; $\Delta \textrm{NLL}_{test} = 0.22 \pm 0.06^*$).

To evaluate the \textbf{cost of defeat} for cases where both mice attacked, we compared two models. In the first model, the cost of defeat was not fixed and remained an optimization parameter. In the second model, the cost of defeat has the same absolute value as the reward for a victory (\cref{fig:model_comparison}H). Our results indicate that the cost of defeat has an absolute value significantly larger than that of the reward for a victory ($\Delta \textrm{NLL}_{train} = 0.61 \pm 0.14^*$; $\Delta \textrm{NLL}_{test} = 0.85 \pm 0.11^*$).

Finally, to evaluate the \textbf{cost of defense} for cases where only one mouse attacked, we considered an alternative model where this cost was equal to zero (\cref{fig:model_comparison}I). We determined that, although the punishment for losing in a fight while “defending” was negligible compared to the punishment for losing in a fight while “attacking”, it still has a non-zero value ($\Delta \textrm{NLL}_{train} = 0.003 \pm 0.003^{ns}$; $\Delta \textrm{NLL}_{test} = 0.019 \pm 0.006^*$). We did not perform an additional ablation study for the large costs of defense, as, in that case, the predicted actions are always to “attack”, unlike in the data (\cref{sup:gametheory_rewardless}).

\subsubsection{Optimal parameters}

To obtain further insights into the aggressive behavior of mice in our experiment, we evaluated the optimal parameters for the first-order Bayesian belief-based model. We observed a local minimum of the NLL at the parameter values $\sigma_1 = 3 \pm 1$g, $\sigma_2 = 6 \pm 2$g, $\beta_o = 5 \pm 1$, $\beta_a = 9 \pm 1$, $\alpha = 0.3 \pm 0.2$, $\mathcal{A} = 3 \pm 1$, and $\varepsilon = 1.0 \pm 0.1$. The identified parameters $\beta_o$ and $\alpha$ correspond to the policy with $\delta = 0$ where mice attack any opponents of equal or lower strength, provided that their action is also to “attack”. The uncertainty $\sigma_1$ in initial estimates of own strength has a relatively low value suggesting that mouse body weight is an informative proxy for its strength. The uncertainty $\sigma_2$ in initial estimates of the opponent’s strength is relatively high suggesting that the opponent’s body weight carries no significant information about its estimated strength and that such strength is rather estimated based on their actions.

\begin{figure}[ht]
\vskip 0.1in
\begin{center}
\centerline{\includegraphics[width=\columnwidth]{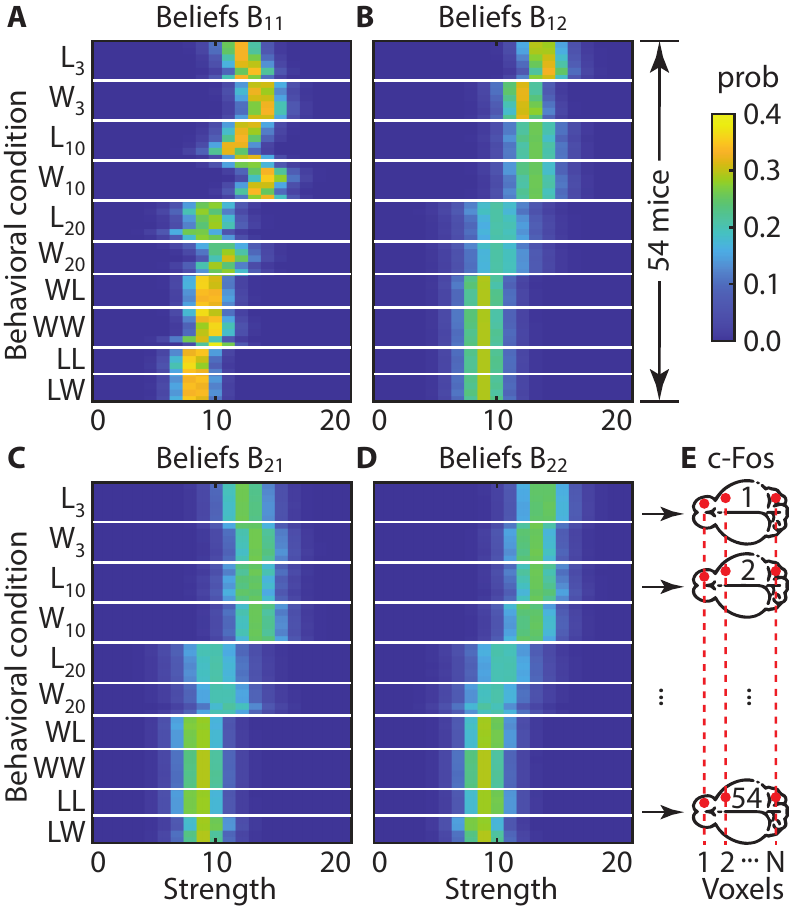}}
\caption{Reconstructed beliefs for individual mice: (A) about oneself; (B) about last opponent; (C) about last opponent's belief about oneself; (D) about last opponent's belief about themselves. (E) Belief regression to c-Fos activity in whole-brain samples.}
\label{fig:reconstructed_beliefs}
\end{center}
\vskip -0.2in
\end{figure}

\subsection{Model correlates in the brain}
\label{res:neural}

To analyze neural correlates of the model variables, based on the behavioral data, we estimated the beliefs for each mouse (\cref{fig:reconstructed_beliefs}A-D). We then computed their correlations, voxel by voxel, with the c-Fos activity in the entire brain (\cref{fig:reconstructed_beliefs}E, \cref{app:belief_correlates}). We evaluated the beliefs at the end of the experiments because c-Fos only allows collecting one brain activity snapshot per animal. To study brain activity at different stages of social conflict, we used mice with varied participation in the experiment (3/10/20/22 days).

\begin{figure}[ht]
\vskip 0.1in
\begin{center}
\centerline{\includegraphics[width=\columnwidth]{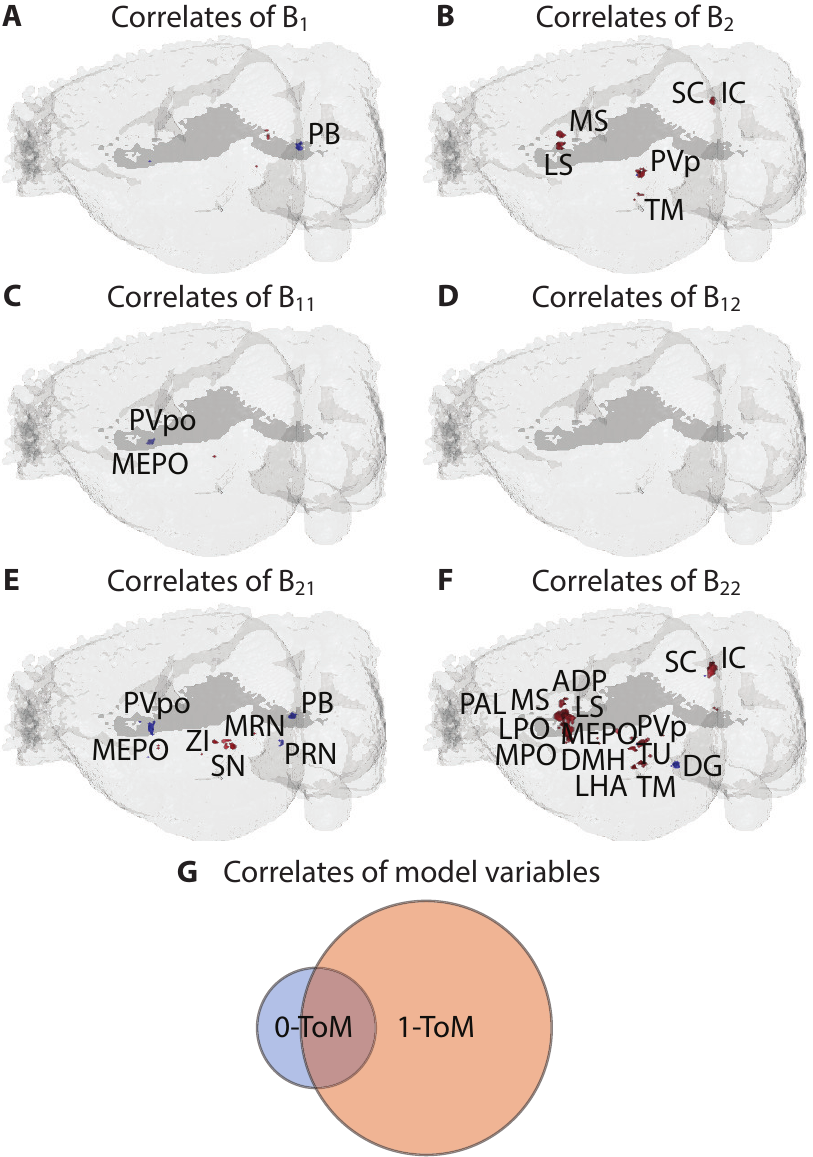}}
\caption{Correlates of the reconstructed beliefs in the c-Fos activity in the brain. (A-B) 0-ToM; (C-F) 1-ToM. Red: positive correlation; blue: negative correlation. (G) Venn diagram for the voxels correlated with the 0-ToM and 1-ToM models.}
\label{fig:belief_correlates}
\end{center}
\vskip -0.2in
\end{figure}

In correlation analysis, we use the ``primary'' and ``secondary'' beliefs reconstructed with 0-ToM and 1-ToM models. We found significant (cumulative FDR $q \leq 0.1$) neural correlates for both types of these variables (\cref{fig:belief_correlates}A-F). To analyze the representations of  0-ToM and 1-ToM beliefs, we examined the set of voxels whose activity was correlated with either 0-ToM or 1-ToM beliefs (\cref{fig:belief_correlates}G), i.e. the union of the voxels correlated with the two models. We found that 79\% of these voxels were correlated with the 1-ToM beliefs only, while 8\% of the voxels were uniquely correlated with the 0-ToM beliefs. The remaining 13\% of voxels were correlated with both 0-ToM and 1-ToM beliefs. Thus, brain activity appears to contain signatures of both 1-ToM and 0-ToM beliefs, suggesting that both models may be relevant to the animals' behavior.  

We then evaluated the brain regions hosting the significant neural correlates of the 1-ToM beliefs. We found that the ``primary'' beliefs about oneself and the opponents were correlated with clusters of neural activity in the median preoptic nucleus (MEPO) and the periventricular hypothalamic nucleus (PV). The ``secondary'' beliefs were correlated with neural activity in the brain regions including, most prominently, the medial septal nucleus (MS), the medial and lateral preoptic areas (MPO, LPO), the lateral septal nucleus (LS), the superior and inferior colliculi (SC, IC), the dentate gyrus (DG), the parabrachial nucleus (PB), the anterodorsal and median preoptic nuclei (ADP, MEPO), the periventricular hypothalamic nucleus and the dorsomedial nucleus of the hypothalamus (PV, DMH), the pallidum (PAL), the zona incerta (ZI), the tuberomammillary and tuberal nuclei (TM, TU), the pontine reticular nucleus (PRN), and the lateral hypothalamic area (LHA). Some of these regions are known for their involvement in conflict behaviors \citep{aleyasin2018, diaz2020}. Our results show that the belief-based model is consistent with neural activity in the brain. We describe additional tests in \cref{sup:additional_correlates} and \cref{fig:additional_correlates}.

\section{Discussion}
\label{res:theory}
\subsection{Normative theory of social conflict}
In this work, we formulated a theory of chronic social conflict based on game theory and Bayesian inference of agents' beliefs. We refined our model using mouse behavioral data and further validated it using neuronal data. 

Our results suggest that animals’ behavior during conflict is consistent with maintaining and updating beliefs about their strength and the strengths of their opponents. This observation is similar to prior work on Bayesian Theories of Mind proposing that in a variety of tasks humans and animals make decisions based on estimates of environmental variables updated with the Bayes rule \cite{baker2011}. This finding also supports prior work showing the lack of naive (without evolving beliefs) game-theory-optimal behaviors in humans \cite{stahl1995}.

Our data suggest that, along with the ``primary'' beliefs about their own and opponent's strengths, animals maintain the ``secondary'' beliefs estimating the ``primary'' beliefs of opponents. This result is consistent with previous work in humans reporting that choices in the volunteer's dilemma are better explained with the 1-ToM model compared to 0-ToM and 2-ToM models \cite{khalvati2019}. Our results are also consistent with behavioral observations in mice suggesting that animals facing a dominant/submissive opponent anticipate their actions \cite{kudryavtseva2014}.

We show that initial estimates of animals' own strength were correlated with their body weight. This finding is consistent with the proposals that body weight is a strong predictor for fighting performance in mice \cite{andersson1996, cooper2020} and social status in rats \cite{nagy2023}. At the same time, our data suggest that animals may not use correlates of opponents' body weights as surrogates for strength, rather relying on opponents' actions.

Our fits to behavioral data suggest that the decisions about animals' actions in the chronic conflict task were made before agonistic interactions with other mice and were not optimized within a single interaction. These fits are consistent with our observations of animal behavior in individual interactions. This result also aligns well with prior work in humans reporting bounded rationality in a series of normal-form games \cite{stahl1995}. Further experiments may be designed to validate this result.

Using behavioral data, we arrived at a payoff matrix describing the animals' decisions in social conflict. We found that the cost of defense (via escaping confrontation) is small compared to the cost of defeat (via attack), in agreement with minimizing the physical damage reported by \cite{crowcroft1966}. At the same time, the cost of defeat after an attack exceeds the reward for victory. This may be due to the immediate effect of the cost of defeat (e.g. physical damage) compared to a discounted delayed reward (e.g. mating, access to food, etc.) associated with winning a conflict \cite{kable2007, kobayashi2008}.

Finally, we identified the neural correlates of the beliefs computed in our model. Previous studies have characterized multiple brain regions whose activity modulates aggression. Our approach offers an interpretation of the computations performed in several brain regions. We find correlates of beliefs in hypothalamic regions, consistent with the previously reported role of the hypothalamus in aggression \citep{aleyasin2018, diaz2020}. Within the hypothalamus, a recent study describes the \textit{short-term} aggression ramping encoded in the ventromedial hypothalamus (VMHvl) but not in the medial preoptic area (MPO) \citep{nair2023}. Here, we observe the correlates of animals' beliefs in preoptic areas (MPO, LPO, MEPO) but not in VMHvl in the conditions of \textit{chronic} social conflict. This finding may indicate differences in the coding schemes for short-term and chronic conflict. We find that the neural activity in MPO is correlated with ``secondary'' beliefs about opponents' strengths, in agreement with a recent study on the role of cMPO neurons \citep{wei2023}. It is likely that other brain regions, such as VMHvl encode variables different from beliefs. This possibility may explain why we did not find significant correlates of beliefs in other brain regions related to aggression including some limbic areas, the brain stem, and the reward system \citep{aleyasin2018}. 

\subsection{Limitations and strengths of our approach}
In our setting, animals made their decisions in the conditions of incomplete information and refined their policies based on observed outcomes of their encounters. Agents in such a setting face the problem of a partially observable Markov decision process (POMDP). Here, we use biological decision-making propose an approach to solving the POMDP problem in multi-agent settings. Our work extends machine-learning methods for analyzing behavioral and neural data in these conditions. Combining inverse rational control \cite{dvijotham2010, kwon2020} with game theory \cite{cressman2003} offers an interpretation of model parameters. Specifically, the variables in our models either represent the rewards for various outcomes or describe the confidence in taking various actions. A low number of fitting parameters makes our approach data-efficient. This property may be useful in neuroscience where data are typically small \cite{stahl1995}. Game theory makes the transitions between beliefs deterministic, reducing the computations compared to probabilistic Baum-Welch algorithms \cite{baum1970}, conventionally used in POMDPs \cite{wu2018}. The benefits of our approach can be quantified in follow-up studies.

A limitation of our game-theory approach is that it involves model comparisons between {\it ad hoc} models \cite{stahl1995}. To mitigate this limitation, we based our models on the results of previous studies, using standard ToM/ML baselines \cite{stahl1995, khalvati2019}. In future studies, we may consider additional models including mixtures of ToMs \cite{khalvati2019}, naive Nash, and rational expectation strategies \cite{stahl1995}. 

Our behavior-based model reconstructs relevant variables for individual animals, allowing us to search for their correlates in neural activity. As a proxy of neural activity, we used the c-Fos expression, captured with whole-brain 3D microscopy \cite{renier2016}. Using c-Fos comes with several limitations: i) c-Fos lacks temporal resolution, allowing one snapshot of brain activity per mouse; ii) the c-Fos signal is nonspecific, potentially encoding factors other than neuronal activities \cite{herrera1996}. Among the benefits, c-Fos does not perturb animals’ social behavior, making it the standard choice for studies of aggression \cite{aleyasin2018, diaz2020, wei2021}. \textit{Ex-vivo} 3D imaging allowed us to reconstruct the whole-brain neural activity. While previous studies focused on tissue sections of preselected brain regions \cite{haller2006, konoshenko2013}, our whole-brain data has enabled an unbiased study of conflict circuitry.

The neural activation analysis in this work is performed with the correlation analysis followed by the FDR correction \cite{benjamini1995}. This is a conservative analysis establishing whether the model variables are directly represented in neural activity. Alternative approaches could include GLMs for detecting linear combinations of model parameters \cite{lindquist2008}, Gaussian random fields for modeling spatial correlations of the signal \cite{rue2005}, and MANOVA for distinguishing representations of correlated model variables \cite{allefeld2014}. Using these and other fMRI approaches \cite{skup2010, ashburner2014} may allow identifying the correlates of individual model variables.

Our results are based on data comprising ``attack''/``defend'' and ``win''/``lose'' readouts but our analyses can be extended to different metrics and social behaviors. Long-term tracking of social structure in rats \cite{nagy2023} may allow for testing our conclusions in a different species in a complex social environment. The data on carrier/non-carrier phenotypes \cite{schroeder2005} would allow us to see if our results generalize to a cooperative setting. The model may also be applied to other domains, such as automated negotiation \cite{baarslag2016}. Overall, our framework may be used for studying competitive or cooperative behaviors in real-world settings using limited data.

\subsection{Broader impact}

In our model, we applied evidence accumulation approaches to the domain of game theory which studies optimal interactions between agents. This allowed us to build a theory of chronic social conflict which can be used in future works for building quantitative models of social interactions. We then used Inverse Rational Control (IRC) to model the beliefs of animals based on their behavior. Combining the IRC with evidence accumulation models limits its degrees of freedom, increases robustness, and offers an interpretation for its predictions. Finally, we used the whole-brain c-Fos data (a proxy of neuronal activity) in combination with the IRC for an unbiased search for neural correlates of reconstructed beliefs. Overall, we combine the game theory, evidence accumulation models, inverse rational control, and whole-brain imaging to propose a principled framework for building normative models of social behaviors and grounding them to neural circuitry in the brain.

\section*{Acknowledgements}
We thank Pavel Osten and Kannan Umadevi Venkataraju for collecting and processing the imaging data; Nikhil Bhattasali and Khristina Samoilova for helpful discussions.

\bibliography{example_paper}
\bibliographystyle{icml2023}

\newpage
\appendix
\twocolumn
\section{Methods}
\subsection{Mouse chronic social conflict paradigm}
\label{app:behavior}
To induce varied behavioral states in mice, we applied the chronic social conflict paradigm (all animal procedures were approved by the Stony Brook University Institutional Animal Care and Use Committee in accordance with the NIH regulations). Pairs of weight-matched mice were separated by a perforated partition in cages. Once daily, the partition was removed for 10 minutes to enable agonistic interactions between mice. To study the adaptive properties of the aggressive/defeated states in mice, we kept the opponents unchanged for 3 days. Then, to reveal the maladaptive properties of behavioral states, each winning mouse was kept in its cage, while each losing mouse was daily relocated to an unfamiliar cage with an unfamiliar winner. Non-fighting mice were matched with known winning opponents. After a total of 20 days of agonistic interactions, to reveal the flexibility of behavioral states, mice were reorganized to face opponents of the same state. The newly formed pairs underwent 2 more days of interactions. During all interactions, we logged the mouse actions (“attack”, “defend”) and the outcomes of interactions (“win”, “lose”, “draw”). We varied the participation of mice in the experiment:
\begin{table}[ht]
\vskip -0.1in
\caption{Datasets}
\label{tbl:datasets}
\begin{center}
\begin{small}
\begin{sc}
\begin{tabular}{lcccr}
\toprule
Dataset & I & II & III & Total \\
\midrule
Days & 3 / 10 & 20 & 22 & \cr \\
\makecell{Mouse \\ behavior} & \makecell{Exp: 33 \\ Ctrl: 6} & \makecell{Exp: 49 \\ Ctrl: 6} & \makecell{Exp: 52 \\ Ctrl: 6} & \makecell{Exp: 134 \\ Ctrl: 18} \cr \\
\makecell{Brain \\ 3D data} & \makecell{Exp: 24 \\ Ctrl: 6} & \makecell{Exp: 11 \\ Ctrl: 6} & \makecell{Exp: 19 \\ Ctrl: 6} & \makecell{Exp: 54 \\ Ctrl: 18} \\
\bottomrule
\end{tabular}
\end{sc}
\end{small}
\end{center}
\vskip -0.1in
\end{table}
\subsection{Game theory optimal actions}
\label{app:gametheory}
We then defined a class of models to test against the logged data. We described the mouse task as a normal-form game. Here, the agents simultaneously chose their actions; the available actions were to “attack” and to “defend”. The outcome of the game depended on the actions and strengths of participating agents. The strengths $s_i$ were defined as constants between 1 -- 20, specific to an agent and unchanging.
\begin{table}[ht]
\vskip -0.1in
\caption{Game outcomes}
\label{tbl:game_outcomes}
\begin{center}
\begin{small}
\begin{sc}
\begin{tabular}{lcccr}
\toprule
Action & \#2 "attacks" & \#2 "defends" \\
\midrule
\#1 "attacks" & $p^1_{win} \sim Gibbs(s_1, s_2)$ & \#1 wins \\
\#1 "defends" & \#2 wins & draw \\
\bottomrule
\end{tabular}
\end{sc}
\end{small}
\end{center}
\vskip -0.1in
\end{table}

If both agents in the game chose to “attack”, the probability of “winning” the game was defined by the softmax rule over their strengths parameterized with the ``outcome confidence'' $\beta_o$. The reward assigned to each agent depended on the outcome of the game and on their action as follows:
\begin{table}[ht]
\vskip -0.1in
\caption{Rewards}
\label{tbl:game_rewards}
\begin{center}
\begin{small}
\begin{sc}
\begin{tabular}{lcccr}
\toprule
Action & "Win" & "Lose" & "Draw"\\
\midrule
"Attack" & 1 & $-\mathcal{A}$ & N/A \\
"Defend" & N/A & $-\alpha$ & 0 \\
\bottomrule
\end{tabular}
\end{sc}
\end{small}
\end{center}
\vskip -0.1in
\end{table}

To derive the optimal actions for agents in a fully observable setting, we parametrized the agents’ actions by the probability to “attack” and derived the gradients of the expected reward w.r.t. these probabilities. We found Nash equilibria corresponding to mixed strategies by deriving the points of zero gradients within the support of action probabilities. To find Nash equilibria corresponding to pure strategies, we chose the pure strategies whose gradients were directed outside the support of action probabilities. We averaged the action probabilities over Nash equilibria. Overall, finding the optimal policies in the game is described in \cref{alg:gametheory}:

\begin{algorithm}[!h]
   \caption{Game theory optimal actions}
   \label{alg:gametheory}
\begin{algorithmic}
   \STATE {\bfseries Input:} strength range $s_{max}$, costs $\alpha$, $\mathcal{A}$, confidence $\beta_o$
   \STATE Initialize actions $A = \textrm{zeros}(2, s_{max}, s_{max})$.
   
   \STATE
   \FOR{strengths $s_1, s_2 = 1$ {\bfseries to} $s_{max}$}
   \STATE expectation $\kappa_1 = (1 + \mathcal{A})\mathop{\textrm{softmax}}_{s}(\beta_o s_1, \beta_o s_2) - \mathcal{A}$
   \STATE gradient $\dot{p}_1 = (\kappa_1 + \alpha - 1) p_2 + 1$; similarly find $\dot{p}_2$
   \STATE
   \STATE Find mixed strategy: $\dot{p}_{1(2)} = 0$; $0 < p_{1(2)} < 1$;
   \FOR{policies $p_1, p_2 = 0$ {\bfseries or} $1$}
   \STATE Find pure strategy: $\dot{p}_{1(2)}(p_{1(2)} - 0.5) > 0$
   \ENDFOR
   \STATE Average the strategies: $A(1, s_1, s_2) = \mathrm{mean}(\{p_1\})$
   \STATE $A(2, s_1, s_2) = 1 - A(1, s_1, s_2)$

   \ENDFOR
\end{algorithmic}
\end{algorithm}

\subsection{Bayesian belief initialization}
\label{app:beliefinit}
To account for the partial observability of information in the task, we defined the probability distributions (beliefs) describing the strengths of the agents. We initialized the beliefs with the normal distributions whose mean values corresponded to the body weight of the animals (shifted by -15g to fit the range). To account for the prior information about the body weight distribution, we applied the Bayes rule to the initial beliefs. To reflect possible depths of reasoning, we considered four types of beliefs parameterized by different standard deviations of the normal distribution:
\begin{table}[ht]
\vskip -0.1in
\caption{Initial belief standard deviations}
\label{tbl:belief_std}
\begin{center}
\begin{small}
\begin{sc}
\begin{tabular}{lcccr}
\toprule
Belief & "Mine" & "Opponent's"\\
\midrule
"about myself" & $\sigma_1$ & $\sigma_1 + \sigma_2$ \\
"about opponent" & $\sigma_2$ & $\sigma_2 + \sigma_1$ \\
\bottomrule
\end{tabular}
\end{sc}
\end{small}
\end{center}
\vskip -0.1in
\end{table}

The ``opponent's'' beliefs here reflect the agent's beliefs about the opponent's beliefs that may differ from the opponent's beliefs in the model. Overall, the acquisition of initial beliefs about the animals' strengths is summarized in \cref{alg:beliefinit} below. Here $B_{m_1 m_2}^{i}$ describes the beliefs of mouse $m_1$ about mouse $m_2$.

\begin{algorithm}[!h]
   \caption{Bayesian belief initialization}
   \label{alg:beliefinit}
\begin{algorithmic}
   \STATE {\bfseries Input:} body weights $w_i$, uncertainties $\sigma_1, \sigma_2$
   \STATE Initialize beliefs $B = \textrm{zeros}(s_{max}, \max(m), \max(m))$.
   \STATE strengths $Pr(s) = P_{\mathcal{N}\left(\mathbb{E}[w], \mathbb{D}[w]\right)}$
   
   \STATE
   \FOR{mice $m_1, m_2 = 1$ {\bfseries to} $\max(\{m\})$}
   \STATE Set $\sigma$ using \cref{tbl:belief_std}
   \FOR{strength $i = 1$ {\bfseries to} $\max(\{s\})$}
   \STATE belief $B_{m_1 m_2}^i = Z^{-1}P_{\mathcal{N}(i, \sigma)}(w_{m_2}) Pr(s)(i)$
   \ENDFOR
   \ENDFOR
\end{algorithmic}
\end{algorithm}

\subsection{Bayesian belief update}
\label{app:beliefupdate}
To enable evidence accumulation in our model, we used the Bayes rule to update the beliefs based on the observations. The agonistic interaction outcomes were used to update the primary beliefs about the agent's own and the opponent's strengths. The opponents' actions were used to update the beliefs about the opponent's beliefs. To speed up computations, we precomputed the conditional probability matrices. Overall, the update procedure for the beliefs about the animals' strengths is summarized in \cref{alg:beliefupdate} below.

\begin{algorithm}[!h]
   \caption{Bayesian belief update}
   \label{alg:beliefupdate}
\begin{algorithmic}
   \STATE {\bfseries Input:} pairs $p_m^t$, actions $a_m^t$, outcomes $o_m^t$
   \STATE Compute actions $A$ using \cref{alg:gametheory}; precompute
   \STATE $[RA]^{a}_{ijkl} \equiv \sum_b R^{ab}_{ij} \textrm{conv}_{kl}(A, P_{\mathcal{N}(0, \sigma_1 + \sigma_2)})^{b}_{kl}$
   \STATE $[OA]^{oab}_{ijkl} \equiv O^{oab}_{ij} \textrm{conv}_{kl}(A, P_{\mathcal{N}(0, \sigma_1 + \sigma_2)})^{b}_{kl}$
   \STATE Initialize beliefs $B$ using \cref{alg:beliefinit}
   
   \STATE
   \FOR{time $t = 1$ {\bfseries to} $\max(\{t\})$}
   \FOR{mouse $m_1 = 1$ {\bfseries to} $\max(\{m\})$}
   \STATE opponent $m_2 = p_{m_1}^t$
   \FOR {mouse {\bfseries in} $m_1, m_2$}
   \STATE $\tilde{a}_k^t = \mathop{\textrm{softmax}}_{a} \sum_{ijkl} [RA]^{a}_{ijkl} B_{11}^i B_{12}^j B_{21}^k B_{22}^l$
   \ENDFOR

   \FOR{strength $i = 1$ {\bfseries to} $\max(\{s\})$}
   \FOR{mice {\bfseries in} $m_1, m_2$}
   \STATE $\Delta B_{12}^i = Z^{-1}\sum_{jkl} [OA]^{o_1 a_1 a_2}_{ijkl} B_{11}^i B_{12}^j B_{21}^k B_{22}^l$
   \STATE update $B_{12}^i \leftarrow \varepsilon \Delta B_{12}^i + (1 - \varepsilon) B_{12}^i$
   \STATE Repeat for other belief types $B_{**}$
   \ENDFOR
   \ENDFOR
   \ENDFOR
   \ENDFOR
\end{algorithmic}
\end{algorithm}

\subsection{Model fit}
\label{app:modelfit}
Besides the belief update, the above algorithm allows for predicting the animals' actions based on their beliefs. We consider the opponent's predicted action to be game-theory-optimal based on the opponent's estimated beliefs. To predict the agent's action, we maximize its reward expectations based on its own ``primary'' beliefs and the opponent's predicted action. To optimize the model predictions, we update the model's parameters to minimize the negative log-likelihood (NLL) for the actions in our behavior data to be sampled from the model predictions. While we use the entire data to propagate the beliefs, we only use days 1-3 and 21-22 for the NLL evaluation to avoid the impact of repeated actions on days 4-20. We use bootstrap to compute the error bounds for the identified parameters. Overall, the inference of model parameters is summarized in \cref{alg:parameterinference} below.

\begin{algorithm}[!h]
   \caption{Model parameter inference}
   \label{alg:parameterinference}
\begin{algorithmic}
   \STATE {\bfseries Input:} pairs $p_m^t$, actions $a_m^t$, outcomes $o_m^t$
   \STATE Initialize parameters $x \equiv [\sigma_1, \sigma_2, \beta_a, \beta_o, \alpha, \mathcal{A}, \varepsilon]$
   \STATE Initialize $x_{opt} = zeros(100, 7)$
   
   \STATE
   \FOR{repeat $i = 1$ {\bfseries to} $100$}
   \STATE Define subset $\tau \in [1, T]$ with repetitions
   \STATE Define actions $Pr(a_m^\tau)$ using \cref{alg:beliefupdate}
   \STATE Define likelihood $\mathcal{L} = -\sum_{mt}\log Pr(a_m^\tau)$
   \STATE Regularize $\mathcal{L} \leftarrow \mathcal{L} + \lambda\left\lVert x/x_{\textrm{max}} \right\rVert_2$
   \STATE Optimize $x_{opt}(i) =$ \textit{fminsearch}($\mathcal{L}$)
   \ENDFOR
   \STATE Compute $\mathbb{E}[x_{opt}], \mathbb{D}[x_{opt}]$
\end{algorithmic}
\end{algorithm}

To optimize the parameter inference procedure, we applied it to simulated data with known parameters. We added noise to \cref{alg:beliefinit} to generate the initial beliefs about strengths in the model and used \cref{alg:beliefupdate} to generate the simulated actions. Once the model parameters were reconstructed with \cref{alg:parameterinference} (no bootstrap), we used Gaussian processes to estimate the means and the error bounds for the predictions. We varied the regularization parameter $\lambda \in \{0.1; 1; 10\}$ and chose the one with the best parameter reconstruction.

\subsection{Model comparison}
\label{app:modelcompare}
To test hypotheses about mouse decision-making, we performed the model comparison. To this end, we optimized pairs of models on the training data (52 mice participating for 22 days) and computed the changes in the NLL based on the predictions for the testing data (82 mice participating for 3/10/20 days). We performed the t-tests on the changes in the NLLs for individual mice followed by the FDR corrections. For pairwise comparisons, we picked an initial model and considered all one-step deviations from it (e.g. for the 0-ToM model, we separately considered “frozen” prior beliefs and, alternatively, a zero cost of defeat, but not their combination). We fitted each model in the same way, using a numerical optimizer until convergence. We considered new models until one of the models outperformed the initial one; we then used this new model as an initial model for the next round of comparisons, using one-step deviations from this new model. We stopped after arriving at a model that outperformed all other models. Here we list the considered models and specify their differences in comparison with the ``default'' (1-ToM) model described above:

\begin{table}[H]
\vskip -0.1in
\caption{Baseline models}
\label{tbl:baseline_models}
\begin{center}
\begin{small}
\begin{sc}
\begin{tabular}{lcccr}
\toprule
Test & Baseline model\\
\midrule
Class of algorithms & Rescorla-Wagner\\
Depth of reasoning & 0-Theory of Mind\\
Flexibility of policy & In-fight Nash equilibria\\
\bottomrule
\end{tabular}
\end{sc}
\end{small}
\end{center}
\vskip -0.1in
\end{table}

In \textbf{Rescorla-Wagner} model, each action (``attack'', ``defend'') was associated with a value representing the expected future reward. This value was independent of the mouse's strength or the identity of an opponent. The values were initialized with zeros for the ``attack'' and ``defend'' actions, then updated at a learning rate $\varepsilon$ (an additional optimization parameter) using the reward as a teaching signal. In the \textbf{0-Theory of Mind} model, we only considered the ``primary'' beliefs of an agent (about its own and the opponent's strength) assuming that the opponent acts in accordance with the same beliefs (even though the actual beliefs of the two agents were independent and may have differed). The actions were selected based on the precomputed Nash equilibria; the opponents' actions and the interaction outcomes were used to update both beliefs. In the \textbf{in-fight Nash equilibria} model, the policy gradients used in determining the Nash equilibria were computed for the expected rewards integrated with the ``primary'' beliefs of each mouse reflecting the in-fight adaptation to the opponent's actions. As the opponent's actions were observable in this setting, the model did not include the ``secondary'' beliefs. 

\begin{table}[ht]
\vskip -0.1in
\caption{Ablation studies}
\label{tbl:ablation_studies}
\begin{center}
\begin{small}
\begin{sc}
\begin{tabular}{lcccr}
\toprule
Test & Ablation\\
\midrule
Dynamics of beliefs & Fixed prior beliefs \\
                    & Fixed posterior beliefs \\
                    & Unit learning rate \\
Role of body weight & Shuffled body weights \\
                    & Wide body weight prior \\
Cost of defeat & Unit cost of defeat\\
Cost of defense & Zero cost of defense\\
\bottomrule
\end{tabular}
\end{sc}
\end{small}
\end{center}
\vskip -0.1in
\end{table}

In the \textbf{fixed prior beliefs} model we used the zero learning rate for the Bayesian belief update, retaining the beliefs unchanged from their body weight-based initial values. In the \textbf{fixed posterior beliefs} model, we used the zero learning rate but the initial beliefs were substituted with the final beliefs produced with our default model. In the \textbf{unit learning rate} model we set the learning rate equal to 1. In the \textbf{shuffled body weights} model, we shuffled the animals' body weights before using them to initialize the beliefs. In the \textbf{wide body weight prior} model we used the double variance for the prior distribution of the animals' body weights at the belief initialization. In the \textbf{unit cost of defeat} model we set the cost of ``losing'' after an ``attack'' to -1, matching the absolute value of the reward for a victory. In the \textbf{zero cost of defense} model we set the cost of ``losing'' after a ``defense'' to 0.

\subsection{Neural correlates of model variables}
\label{app:belief_correlates}
To find neural correlates of the model variables, we normalized the registered c-Fos activity in each brain sample by dividing it by the average c-Fos activity of all control samples from the same dataset. We converted the relative c-Fos activities to the logscale to obtain a roughly normal distribution of relative activities. We normalized the model variables by their average estimated values for the control samples. Then, we computed the voxel-wise correlations between the processed c-Fos activity and model variables. To select the voxels significantly correlated with model variables, we performed the FDR correction. Finding the belief correlates in the brain is described in \cref{alg:belief_correlates} below.
\begin{algorithm}[!h]
   \caption{Belief correlates in the brain}
   \label{alg:belief_correlates}
\begin{algorithmic}
   \STATE {\bfseries Input:} 3D brain images $I$ aligned to atlas $A$; beliefs $B$

   \STATE
   \FOR{sample $s$ {\bfseries in} experimental samples}
   \STATE Normalize $I_s = \log{I_s} - \log{(\mathrm{mean}(I_{ctrl}))}$
   \STATE Normalize $b_s = \mathrm{mean}(b_s) - \mathrm{mean}(b_{ctrl})$
   \ENDFOR
   
   \STATE
   \FOR{belief $b$ {\bfseries in} \{$B_{11}, B_{12},$ etc.\}}
   \FOR{voxel $i$ {\bfseries in} 3D brain atlas volume}
   \STATE Compute $p_i \leftarrow \mathrm{corr}(\{I\}_i, b)$
   \ENDFOR
   \STATE Compute $\{q_i\} = \mathrm{FDR}(\cup_i \{p_i\})$
   \STATE Find $corr_b = \{j \leq i\}: \{\sum_{k=0}^{i}{\mathrm{sort}(q_k)} < 0.1\}$
   \STATE Find brain regions $r$ in atlas $A$: $corr_b \in r$
   \ENDFOR
\end{algorithmic}
\end{algorithm}

The variables under consideration included: ``primary'' and ``secondary'' beliefs about mouse's own and the opponent's strength for 0-ToM and 1-ToM models; the estimated advantages (i.e. the differences between the same-order beliefs within each model); the estimated outcomes (i.e. the binarized estimated advantages); the predicted actions for oneself and the opponent under each model; the Rescorla-Wagner values of the two actions and their difference. The control variables included: the body weights of mice, the observed actions, the outcomes of the agonistic interactions, and the shuffled beliefs of each type.

\twocolumn
\section{Supplementary results}
\subsection{Game theory optimal actions}
\subsubsection{General case}
\label{sup:gametheory}
In this section, we provide a simple derivation of the game theory optimal actions in our model. The expected reward $\mathbb{E}[r_1]$ for the first agent can be computed as its payoff matrix $\hat{R}_1$ multiplied by the action probability vectors $P_{1,2}$ describing the policies of the first and the second agents:
\begin{equation}
\begin{gathered}
\mathbb{E}[r_1] = P_1^T \hat{R}_1 P_2 = \\
= \begin{pmatrix} p_1 & 1 - p_1 \end{pmatrix} \begin{pmatrix} \kappa_1 & 1 \\ -\alpha & 0 \end{pmatrix} \begin{pmatrix} p_2 \\ 1 - p_2 \end{pmatrix} = \\
= p_1 p_2 (\kappa_1 + \alpha -1) + p_1 - \alpha p_2.
\end{gathered}
\end{equation}
Here $p_{1,2}$ are the probabilities to ``attack'' for the first and the second agents, $\kappa_1$ is the expected reward for the first agent in the case when both agents ``attack'', and $-\alpha$ is the cost for the first agent in the case when only the second agent ``attacks''. The first agent may increase its expected reward by changing its policy $P_1$ that depends on a single parameter $p_1$ (the first agent can't change the parameter $p_2$ as it is controlled by the opponent). To optimize the expected reward $\mathbb{E}[r_1]$, we compute its gradient w.r.t. $p_1$:
\begin{equation}
\label{eq:reward_gradient}
\frac{\partial \mathbb{E}[r_1]}{\partial p_1} = p_2 (\kappa_1 + \alpha -1) + 1.
\end{equation}
We then set this gradient to zero. The resulting condition for optimizing the first agent's expected reward $\mathbb{E}[r_1]$ is determined by the second agent's probability to ``attack'':
\begin{equation}
p_2 = -(\kappa_1 + \alpha -1)^{-1}.
\end{equation}
Even though this result may seem counterintuitive (the first agent can only optimize $p_1$ but not $p_2$), there is no controversy. That is, both agents are expected to optimize their policies simultaneously. Similarly to the first agent's reward dependence on $p_2$, the second agent's reward depends on $p_1$. As a result, both their rewards are optimized when:
\begin{equation}
\label{eq:mixed_strategy}
\begin{cases}
p_1^{mixed} = -(\kappa_2 + \alpha -1)^{-1}; \\
p_2^{mixed} = -(\kappa_1 + \alpha -1)^{-1}.
\end{cases}
\end{equation}
Regardless of such optimum being a minimum or a maximum of either expected reward $\mathbb{E}[r_{1, 2}]$, it corresponds to a Nash equilibrium as none of the agents can gain an advantage by only changing its strategy. A solution of this kind (``mixed strategy''), however, only exists if $0 \leq -(\kappa_{1, 2} + \alpha -1)^{-1} \leq 1$, as the probabilities to ``attack'' $p_{1,2}$ cannot be negative and cannot exceed one. Alternatively, the agents may follow a ``pure strategy'' with binary probabilities $p_{1, 2}$ to ``attack''. Such strategies correspond to Nash equilibria if the reward gradients point outwards the $[0 - 1]$ interval for both agents so that the policy won't have room to evolve. This requirement can be formalized as:
\begin{equation}
\label{eq:pure_strategy_general}
\begin{cases}
\frac{\partial \mathbb{E}[r_1]}{\partial p_1}(p_1 - \frac{1}{2}) > 0; \\
\frac{\partial \mathbb{E}[r_2]}{\partial p_2}(p_2 - \frac{1}{2}) > 0.
\end{cases}
\end{equation}
Using the values of the derivatives from Equation \ref{eq:reward_gradient}, we arrive at the criterion for the pure-strategy Nash equilibrium:
\begin{equation}
\label{eq:pure_strategy}
\begin{cases}
(p_2^{pure} (\kappa_1 + \alpha -1) + 1)(p_1^{pure} - \frac{1}{2}) > 0; \\
(p_1^{pure} (\kappa_2 + \alpha -1) + 1)(p_2^{pure} - \frac{1}{2}) > 0.
\end{cases}
\end{equation}
Each task setting (defined by the parameters $\kappa$ and $\alpha$) may correspond to multiple Nash equilibria. As it is not known what equilibrium the agents may end up in under an arbitrary set of parameters, we approximate the expected policy by averaging the action probabilities $p_{1,2}$ over the identified equilibria. A more precise way to account for multiple Nash equilibria is to integrate the action probabilities over the basins of attraction corresponding to each equilibrium; this analysis can be added in future studies.

\subsubsection{Limit case: rewardless victory}
\label{sup:gametheory_rewardless}
In the previous subsection, we set the reward for a victory equal to one, so that we could scale all other rewards in the payoff matrix $\hat{R}$ relative to its value. Such a definition necessitates explicitly considering a limit case where there's no reward for a victory. In this case, the expected reward $\mathbb{E}[r_1]$ for the first agent can be computed as:
\begin{equation}
\begin{gathered}
\mathbb{E}[r_1] = P_1^T \hat{R}_1 P_2 = \\
= \begin{pmatrix} p_1 & 1 - p_1 \end{pmatrix} \begin{pmatrix} \kappa_1 & 0 \\ -\alpha & 0 \end{pmatrix} \begin{pmatrix} p_2 \\ 1 - p_2 \end{pmatrix} = \\
= p_1 p_2 (\kappa_1 + \alpha) - \alpha p_2.
\end{gathered}
\end{equation}
The derivatives of the expected rewards w.r.t. the agents' policies are then defined by:
\begin{equation}
\begin{cases}
\frac{\partial \mathbb{E}[r_1]}{\partial p_1} = p_2 (\kappa_1 + \alpha); \\
\frac{\partial \mathbb{E}[r_2]}{\partial p_2} = p_1 (\kappa_2 + \alpha).
\end{cases}
\end{equation}
Generally, this system only has a trivial solution ($p_1 = p_2 = 0$), ruling out ``mixed'' strategies in this case. The condition for ``pure'' strategies (Equation \ref{eq:pure_strategy_general}) can be written as:
\begin{equation}
\begin{cases}
p_2^{pure} (\kappa_1 + \alpha)(p_1^{pure} - \frac{1}{2}) > 0; \\
p_1^{pure} (\kappa_2 + \alpha)(p_2^{pure} - \frac{1}{2}) > 0.
\end{cases}
\end{equation}
As there is no reward for victory and the parameter $\alpha$ is negligible, the term $(\kappa_{1,2} + \alpha)$ is generally negative. To satisfy the inequality above, the terms $(p_{1,2}^{pure} - 1/2)$ should also be negative, which is only possible with $p_1 = p_2 = 0$ (as we only consider binary values here). That is, the only Nash equilibrium for this limit case is provided by the ``pure'' strategy where none of the agents ever ``attack'', which is inconsistent with the experimental data.

\subsubsection{In-fight Nash equilibrium}
\label{sup:gametheory_flex}
In the case where agents do not have perfect information about each other's strengths, their estimates of the Nash equilibria may be based on their beliefs about the strengths of the agents in the game. As these beliefs, generally speaking, are different between the contestants, their estimates of the Nash equilibria may also be mismatched. As such, these estimates cannot be considered as true Nash equilibria, requiring further evolution of the contestants' policies.

One way to account for this discrepancy is to update the policies during the fight. Although our tests suggest that such a mechanism is inconsistent with animals' behavior in the experiment and, as such, \textbf{is not used in our final model}, we provide its description for completeness.

Updating the policies during the fight is supposed to lead to a joint Nash equilibrium. This involves two simple additions to our previously described model: i) the reward expectations in the payoff matrices $\hat{R}_{1, 2}$ are integrated over the ``primary'' beliefs about the participants' strengths, and ii) the optimization is performed jointly for both participating agents as their policies are observable to each other under the assumptions about this case.

Amending the payoff matrix $\hat{R}_1$ can be reduced to redefining its entry $\kappa_2$, the only one depending on the agents' strengths. Note that an in-fight Nash equilibrium is only defined under the 0-ToM model: the ``primary'' beliefs are used to predict the odds of winning while the ``secondary'' beliefs, normally used for predicting the opponent's actions, are unnecessary because these actions are directly observed:
\begin{equation}
\kappa_2^2 = \sum_{ij} \kappa_2(s_2 = i, s_1 = j) B_{22}^i B_{21}^j.
\end{equation}
That is, the equilibrium policy $p_1$ for the first agent depends on the second agent's estimate of its expected reward $\kappa_2^2$ for the case when both agents ``attack''. The expressions for the optimal ``mixed'' and ``pure'' strategies are then rewritten based on Equations \ref{eq:mixed_strategy} and \ref{eq:pure_strategy} as follows:
\begin{equation}
\begin{cases}
p_1^{mixed*} = -(\kappa_2^2 + \alpha -1)^{-1}; \\
p_2^{mixed*} = -(\kappa_1^1 + \alpha -1)^{-1};
\end{cases}
\end{equation}
\begin{equation}
\begin{cases}
(p_2^{pure*} (\kappa_1^1 + \alpha -1) + 1)(p_1^{pure*} - \frac{1}{2}) > 0; \\
(p_1^{pure*} (\kappa_2^2 + \alpha -1) + 1)(p_2^{pure*} - \frac{1}{2}) > 0.
\end{cases}
\end{equation}
The equations above describe the in-fight Nash equilibrium which is \textbf{not a part of our model}. This analysis, however, may be useful in future studies of different social behaviors.

\subsection{Model fit}
\label{sup:modelfit}
\begin{figure}[ht]
\vskip 0.1in
\begin{center}
\centerline{\includegraphics[width=\columnwidth]{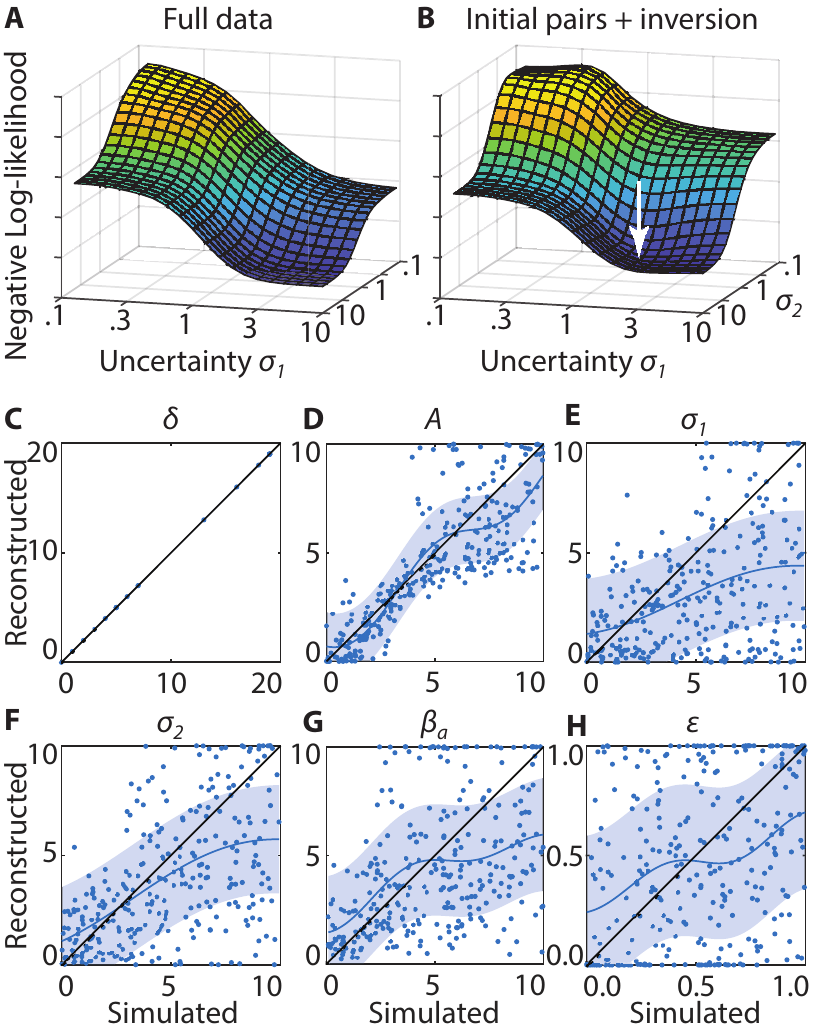}}
\caption{Parameter inference. (A-B) NLL for reconstructing the animal data; (C-H) reconstructed parameters in simulation.}
\label{fig:parameter_inference}
\end{center}
\vskip -0.2in
\end{figure}

In data fits, we used the entire data to propagate the beliefs, however, we only considered days 1-3 and 21-22 for the NLL evaluation. This is because, on days 4-20, winning mice were always matched with losing mice, so the results of the interactions were fully predictable, reducing the sensitivity of the NLL to the model parameters. As a result, the model using days 1-3 and 21-22 had a clear local minimum (\cref{fig:parameter_inference}B) at the identified parameters, whereas the model using days 1-22 had a plateau instead (\cref{fig:parameter_inference}A).

To assess the reliability of our data fits we performed additional fits on simulated data. To this end, we replicated the mouse experiment in a simulated environment and generated 300 choice patterns using various predefined parameters. We sampled parameters uniformly from the ranges previously used for parameter fits in mouse data. We used Gaussian process regression with the squared exponential kernel to estimate the ranges where the reconstruction was accurate (\cref{fig:parameter_inference}C-H). Due to unconstrained degrees of freedom in parameters $\beta_o$ and $\alpha$ we instead computed the parameter $\delta$ determined by $\beta_o$ and $\alpha$ (\cref{fig:gametheory}D). The parameter $\delta$ was reconstructed precisely in all cases. The estimates for the parameter $\mathcal{A}$ were consistent with the data for $0 < \mathcal{A} < 10$. For $\sigma_1$ and $\sigma_2$ the estimates were reliable in the ranges $0 < \sigma_1 < 4$ and $0 < \sigma_2 < 7$, and for $\beta_a$ in the range $0 < \beta_a < 5$. The estimates for $\varepsilon$ were less reliable, although the right trend was observed throughout the range of parameters. The parameters $\delta$, $A$, and $\sigma_{1,2}$, obtained from the mouse data, belong to the above ranges, which renders them accurate. The parameter $\beta_a$ lies outside this range suggesting high confidence in the action.

\subsection{Model comparison}
\label{sup:model_comparison}
\begin{figure}[h]
\vskip 0.1in
\begin{center}
\centerline{\includegraphics[width=\columnwidth]{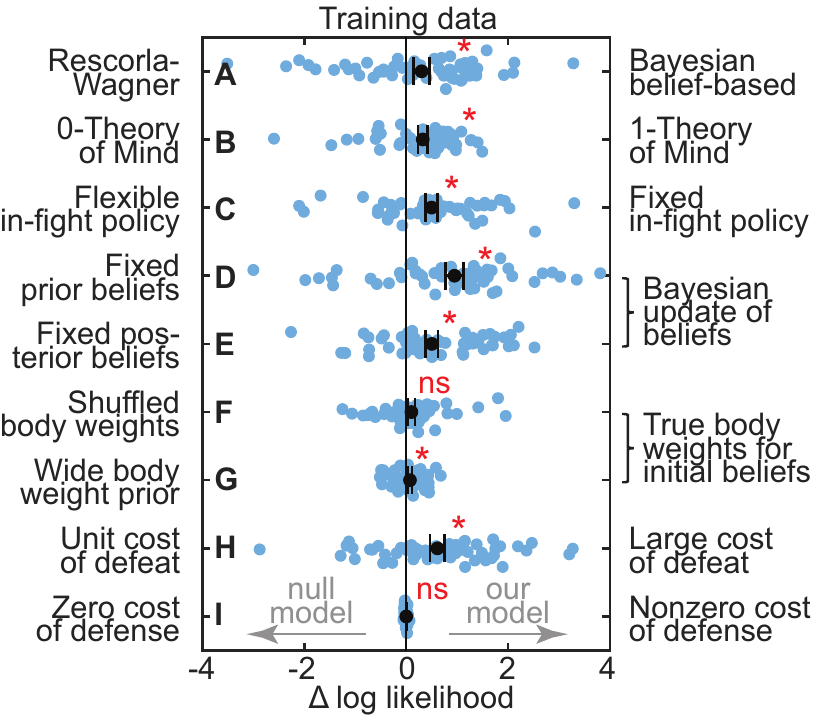}}
\caption{Model comparison on training data. Here (*) indicates a significant difference between the models (t-test $p \leq 0.05$), (ns) indicates a non-significant difference (t-test $p > 0.05$), and the whiskers show the mean $\pm$ the standard error of the mean.}
\label{fig:model_comparison_training}
\end{center}
\vskip -0.2in
\end{figure}

In Results, we provided \cref{fig:model_comparison} displaying the model comparison for the \textbf{testing} data. Here, we provide an additional \cref{fig:model_comparison_training} with a similar display for the \textbf{training} data. In the majority of tests, our model is again more consistent with the data compared to the null models. Even though this data was used to fit the models, the result is still valuable because it covers additional behavioral states: while our testing data covers normal (3 days) and pathological (up to 20 days) conflict, the training data also includes the reversal of the aggressive/submissive status (22 days).

We do not expect overfitting on the training data, as the low number of parameters is low in our approach. Our ToM models contain 7 parameters: (1-2) uncertainties in own and opponent’s strength $\sigma_1$ and $\sigma_2$; (3-4) confidence in outcome and action $\beta_o$ and $\beta_a$; (5-6) punishment for losing while attacking or defending $\mathcal{A}$ and $\alpha$; (7) learning rate $\varepsilon$. Ablations exclude one parameter at a time. The Rescorla-Wagner model does not have belief-related parameters (1-3).

\subsection{Model correlates in the brain}
\label{sup:additional_correlates}
To analyze the \textbf{neural representations} corresponding to the considered models, we compared the c-Fos signal to alternative ways of encoding the belief-related variables. For both 0-ToM and 1-ToM models, we analyzed the correlates of the estimated advantages (i.e. the differences between the same-order beliefs within a model), the estimated outcomes (i.e. the binarized estimated advantages), and the predicted actions for oneself and the opponent. We found that the neural correlates of the beliefs we report in the Results are not explained by these other representations.

\begin{figure}[h]
\vskip 0.1in
\begin{center}
\centerline{\includegraphics[width=\columnwidth]{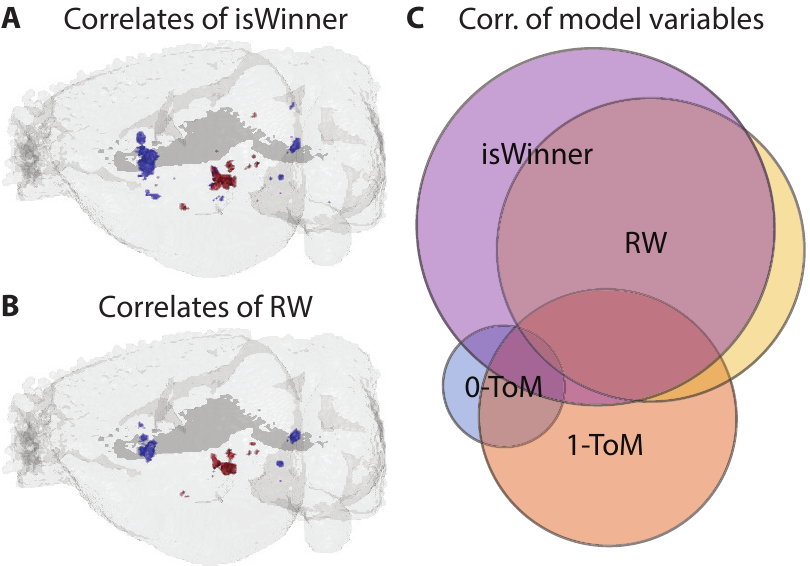}}
\caption{Correlates of the reconstructed beliefs in the c-Fos activity in the brain. (A) winner/loser; (B) Rescorla-Wagner. Red: positive correlation; blue: negative correlation. (C) Venn diagram for the voxels correlated with model variables.}
\label{fig:additional_correlates}
\end{center}
\vskip -0.2in
\end{figure}

To control for \textbf{alternative explanations} of the observed neural activity, we repeated the analysis for model-unrelated variables. These variables included the body weights of the animals (used to initialize the beliefs), the outcomes of the agonistic interactions, and the binary winner/loser variable. As a control, we also considered the shuffled beliefs of each type, such that we shuffled the animal identities within all groups of ``winners'' or ``losers''. Among these, we only found substantial correlates of the winner/loser variable (\cref{fig:additional_correlates}A) whose presence in the brain is unsurprising. The correlates of this variable only partially overlap with the correlates of the beliefs (\cref{fig:additional_correlates}C).

To verify the \textbf{class of algorithms} involved in the animals' aggressive decisions, we additionally examined the neural correlates of variables reconstructed with the Rescorla-Wagner model (\cref{fig:additional_correlates}B). We found that such correlates are mostly explained by the winner/loser variable, whereas a larger portion of the 1-ToM correlates does not have an alternative explanation (\cref{fig:additional_correlates}C). This result corroborates that the animal choices are consistent with the belief-based model at the level of neuronal activity.

\end{document}